\documentclass[11pt]{article}
\usepackage{moriond,epsfig}

\def\PRD{{\em Phys. Rev.} D}

\def\ra{\rightarrow}

\def\be{\begin{equation}}
\def\ee{\end{equation}}
\def\bea{\begin{eqnarray}}
\def\eea{\end{eqnarray}}

\def\met{\not\!\!E_T}
\begin{document}
\vspace*{4cm}
\title{TOP PAIR PRODUCTION CROSS-SECTION AT THE TEVATRON}

\author{ URSULA BASSLER \\
{\it in behalf of the CDF and D\O\ collaboration}}

\address{LPNHE, 4 place Jussieu,\\
75252 Paris Cedex 05, France}

\maketitle\abstracts{An overview of latest top quark pair production
cross-sections measured at the Tevatron is given. These measurements have
been carried out in the dilepton, lepton+jets and all-jets channels
with an integrated luminosity of about $1~\rm{fb^{-1}}$. The measurements
are consistent with NNLO calculations.
}

Since the top quark discovery in $1995$ by the CDF and D\O\ collaborations~\cite{topdisc},
top pair production cross-sections are one of the
basic measurements to be carried out on each new data sample.
During Tevatron Run I (1992-1996) an integrated luminosity of about
$100~\rm{pb^{-1}}$ at a center of mass energy of 
$\sqrt{s}=1.8~\rm{TeV}$ allowed to measure top pair production cross-sections
of $6.5^{+1.7}_{-1.4}~\rm{pb}$ and $5.7\pm 1.6~\rm{pb}$ 
by the CDF and D\O\ collaborations respectively.
The Tevatron Run II started in 2001 and until 
spring 2006 about $1~\rm{fb^{-1}}$ of  $p\bar{p}$ collisions with 
$\sqrt{s}=1.96~\rm{TeV}$ have been produced and analyzed
since. At this energy an increase of about $30\%$ in the cross-section
is expected. The most recent
NNLO calculations predict a cross-section of $6.7^{+0.7}_{-0.9}~\rm{pb}$~\cite{cacciari} or 
$6.8\pm 0.6~{\rm pb}$~\cite{kidonakis} for a top mass, 
$\rm{m_{\rm top}}=175~\rm{GeV}$.

In the standard model (SM) $|V_{tb}| \sim 1$ leads to a branching fraction
$t \ra W b$ close to $100\%$. With a lifetime 
of $\tau \sim 10^{-25}~\rm{s}$ top quarks
decay before hadronization. Their decay channels are classified
according to the decay of the $W$ bosons produced.
The $dilepton$ channel accounts for about $6\%$ of all decays,
taking into account decays into $ee$, $\mu\mu$ and $e\mu$ and including 
leptonic $\tau$ decays.
The $l$+jets channel represents about $34\%$ of the cross-section
with the leptonic $\tau$ decays included in the $e$+jets and $\mu$+jets 
channels. Decays into $all~jets$ occur in $46\%$ of the events, the remaining
$14\%$ correspond to signatures with hadronic $\tau$ decays.

Cross-section measurements are an  important test of perturbative QCD 
at high $p_T$ as non-SM top production, for example 
resonant top production, may lead to higher a cross-section than expected.
It is important to verify the consistency of different decay channels, as
some non-SM models for example $t \ra H^+$ or $t \ra \tilde{t}$, 
modify the contributions in different decay channels.
Non-$W$ top decays are probed 
from the comparison of the $dilepton$ and $l$+jets measurements.
Top quark event selections using $b$-jet tagging assume a branching
ratio $BR(t \ra Wb ) = 1$. Their consistency with kinematic methods,
free of this assumption, is an important check of the SM prediction.

A cross-section is in most cases obtained from a counting experiment:
$ \sigma(p\bar{p} \ra t\bar{t}) =  (N_{\rm obs} - 
N_{\rm bkgd}) / A_{\rm{tot}} {\cal L}$.
$N_{\rm bkgd}$, the number of background events, 
estimated from Monte Carlo simulations and/or
data samples, is subtracted from $N_{\rm obs}$, the 
number of observed events meeting the selection criteria 
of a top-event signature. This difference is
normalized by the integrated luminosity ${\cal L}$ and the total acceptance
$A_{\rm tot}$.
$A_{\rm tot}$ includes the geometric acceptance as well as trigger
efficiency and event selection efficiency and is slightly dependent 
on $m_{\rm top}$. In all the Monte Carlo simulations 
 $m_{\rm top} =175~\rm{GeV}$ has been used.

\section{The dilepton channels}
The signature of top dilepton events is two high $p_T$, opposite sign leptons ($p_T>15~\rm{GeV}$), 
some missing transverse energy ($\met > 35~\rm{GeV}$) and two or more, 
high $p_T$ jets ($p_T>20~\rm{GeV}$). 
Physics background is due dominantly to $Z/\gamma^{\ast}$+jets events,
$WW/WZ/ZZ$+jets events, and estimated from Monte Carlo simulations.
Instrumental backgrounds occur due to fake isolated leptons, 
either as a mis-identified $e$ or a $\mu$ in a non-reconstructed $b$-jet,
as well as $\met$ from detector resolution, fake jets or noise in the
calorimeter. These background are estimated from data.


D\O\ measured the cross-section in the $ee$, $\mu\mu$ and $e\mu$ channels~\cite{d0_dilepton}.
Requiring $2~leptons$ and $2~jets$ in the event selection yields to an acceptance
of $8\%$ and $5\%$ in the $ee$ and $\mu\mu$ channels, and $12\%$ in the $e\mu$ channel.
The acceptance in $e\mu$ channel could be further improved by an additional 
$3\%$ taking in account the events with only $1~jet$ in the final state.
In total $73$ events are observed for $51$ expected signal events and 
$24$ expected background events.
Details for each channel are given in table~\ref{tab:dilepton}.
The combined result from the three measurements is
$\sigma_{t\bar{t}} = 6.8^{+1.2}_{-1.1}{\rm(stat)}^{+0.9}_{-0.8}{\rm(syst)}\pm 0.4{\rm(lumi)~pb}$, 
with the main systematic errors being the lepton identification efficiency and the jet energy
calibration. 
\begin{table}[t]
\caption{Summary of the observed and expected number of events in the dilepton channels used
for the D\O\ combined dilepton cross-section measurement and the CDF $lepton+track$ measurement. \label{tab:dilepton}}
\begin{center}
\small
\begin{tabular}{|l||c|c|c|c||c|}
\hline
 & $ee$ & $\mu\mu$ & $e\mu(\ge 2 jets)$ & $e \mu (1 jet)$ & $l+track$\\
\hline
\hline
${\cal L}\rm{\ in\ fb^{-1}}$ &  $1.04$ & $1.05$ & $1.05$ & $1.05$ & $1.07$ \\
\hline
\hline
$N_{\rm bkgd}$ &  $3.0 \pm 0.5$ & $3.6 \pm 0.5$ & $6.7 \pm 1.2$ & $10.2^{+1.8}_{-1.7}$ & $48.2\pm 4.4$ \\
\hline
$t\bar{t}_{\rm exp}$ &  $9.5 \pm 1.4$ & $5.8 \pm 0.5$ & $28.6^{+2.1}_{-2.4}$ & $7.1^{+0.6}_{-0.7}$ & $60.5 \pm 1.9$\\
\hline
\hline
$N_{\rm obs}$ & $16$ & $9$ & $32$ & $16$ & 129 \\
\hline
\end{tabular}
\end{center}
\end{table}

An alternative method used by CDF loosens the lepton identification criteria for the
second lepton, by requiring only an isolated track~\cite{cdf_lepton_track}. The error
on this measurement has been improved through an increase of the acceptance reaching $14\%$, 
even though the signal/background (S/B) ratio is reduced to $1.3$. The number of expected 
and observed events are also given in table~\ref{tab:dilepton}, leading to a cross-section of
$\sigma_{t\bar{t}} = 9.0 \pm{1.3}{\rm(stat)}\pm{0.5}{\rm(syst)}\pm 0.5{\rm(lumi)~pb}$.

D\O\ carried out an exclusive lepton+track analysis on a data sample with $360~\rm{pb^{-1}}$,
explicitly vetoing a fully reconstructed second lepton to allow for a combination
with the dilepton measurements and using $b$-jet tagging to improve
the purity of the sample. An update of this measurement with $1~\rm{fb^{-1}}$ is in progress.

\section{The lepton+jets channel}

For top decays into $l$+jets, the signature is a high $p_T$ lepton, 
large $\met$ and $4$ or more high $p_T$ jets.
Dominating physics background is due to $W$+jets events. 
Instrumental background is due to fake isolated
leptons in multijet events. 
To separate the signal and background in the $l$+jets channels either the
kinematic properties of the events are used or $b$-jet tagging is required.

\begin{figure}
\begin{center}
\epsfig{figure=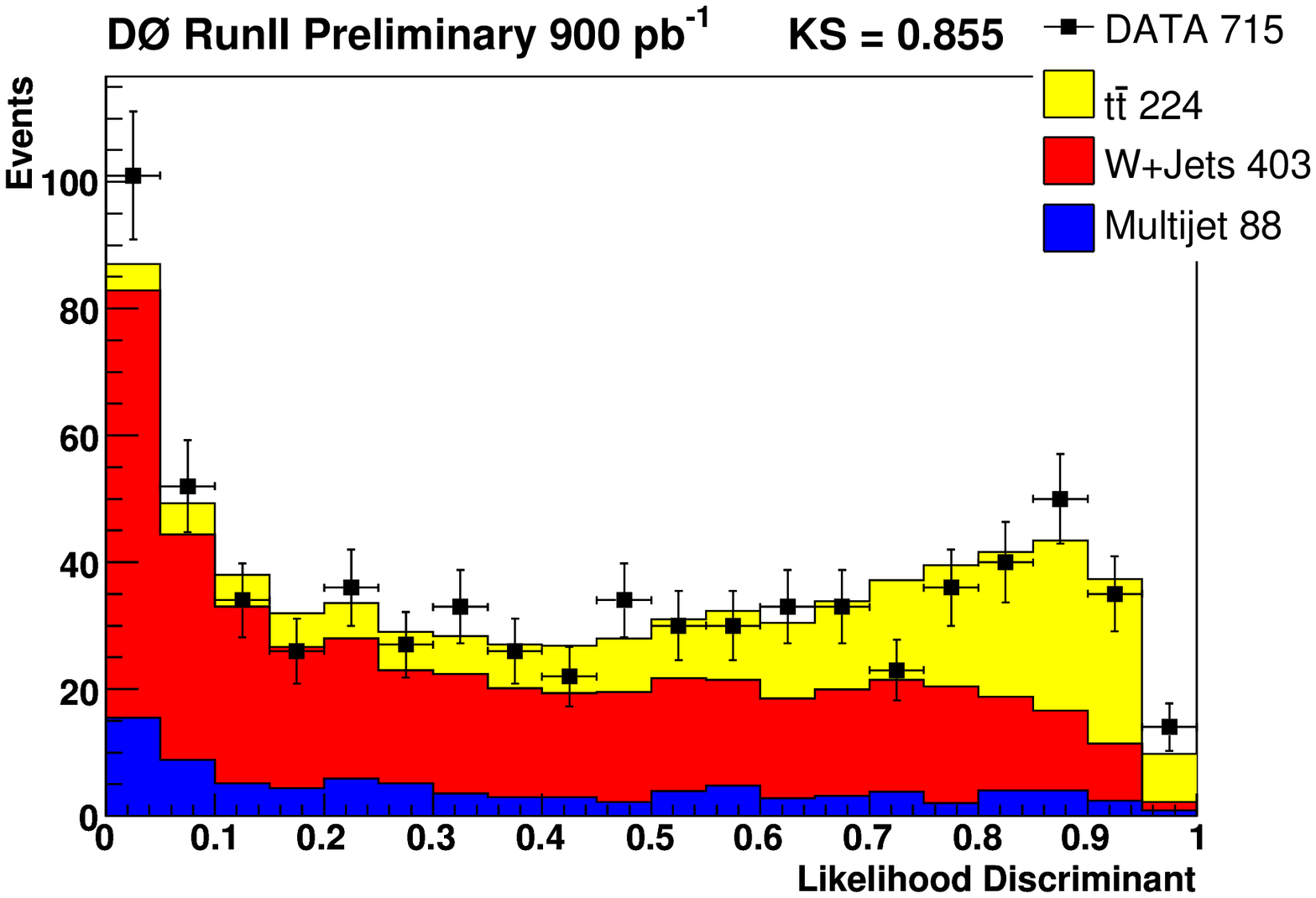,height=3.5cm}
\epsfig{figure=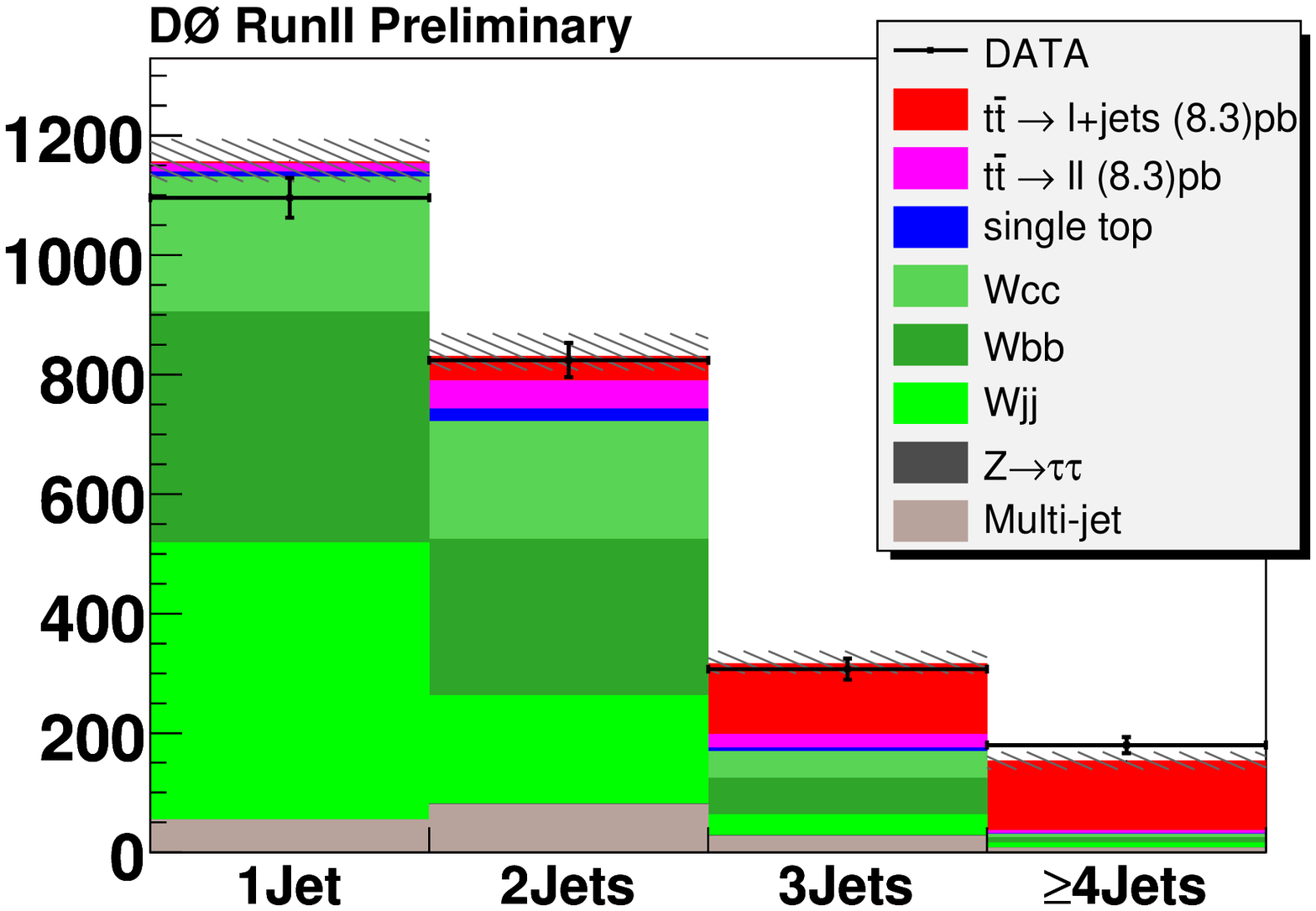,height=3.7cm, clip=}
\epsfig{figure=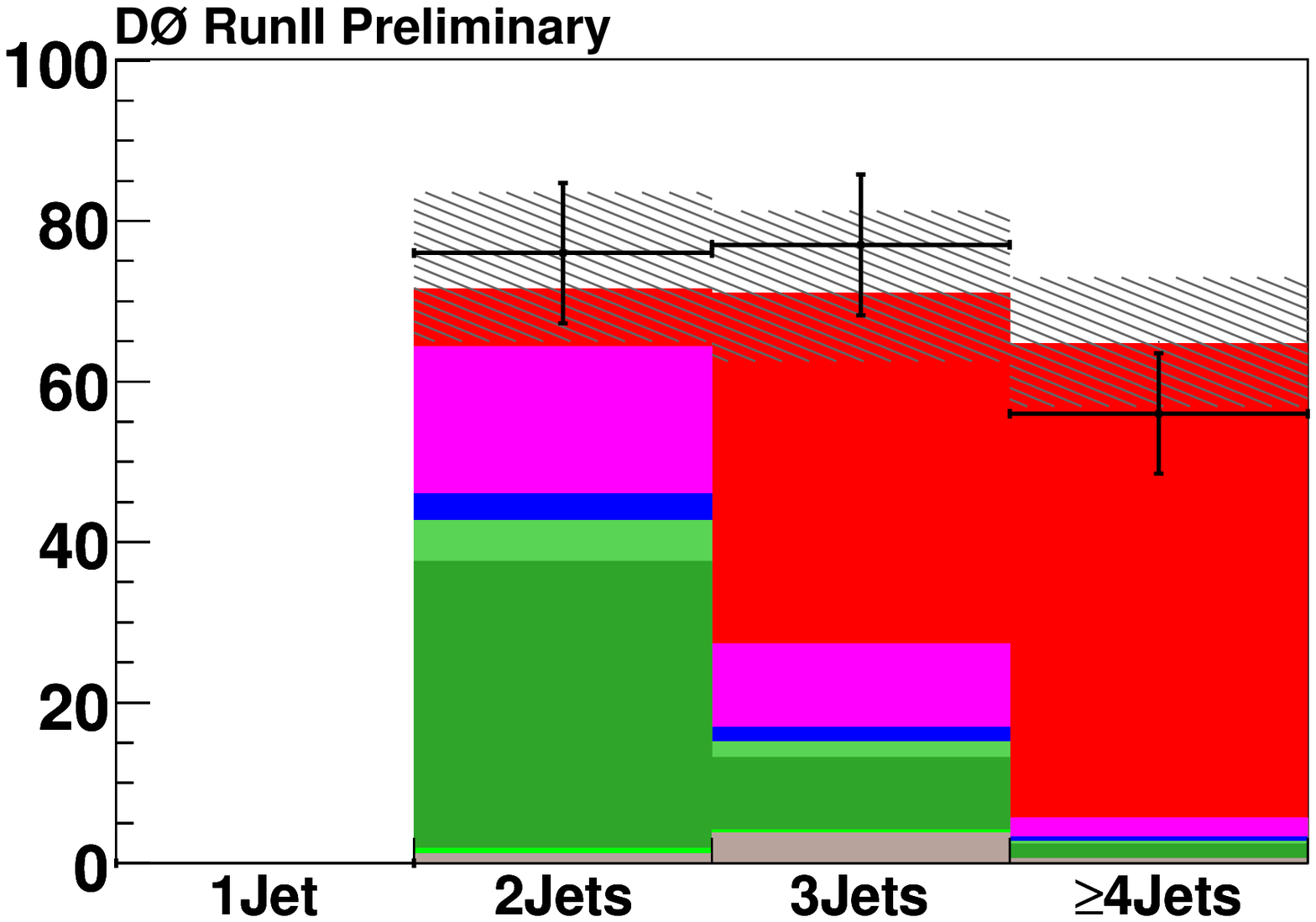,height=3.7cm,clip=}
\caption{The Likelihood discriminant for the kinematic analysis (left) and the jet multiplicities
in the $1$-tag sample (middle) and $2$-tag sample (right) for the $b$-jet tagging analysis with the 
D\O\ lepton+jets samples.
\label{fig:ljets}}
\end{center}
\end{figure}

For the first type of analysis, D\O\ constructs a likelihood discriminant
based on six kinematic variables without a $b$-jet tagging 
requirement~\cite{d0_lepton_jet_kinematic}.
Its output is shown in figure~\ref{fig:ljets}(left) for the combined $e$+jets and $\mu$+jets sample. 
For an integrated luminosity of ${\cal L} =0.91~\rm{fb^{-1}}$, 
$124$ ($100$) $t\bar{t}$ events are expected in the $e$+jets ($\mu$+jets) channel 
for $168$ ($235$) $W$+jets events and $62$ ($27$) multijet events, 
leading to a combined cross-section of
$\sigma_{t\bar{t}} = 6.8^{+0.9}_{-0.8}{\rm(stat)}\pm{+0.7}{\rm(syst)}\pm 0.4{\rm(lumi)~pb}$.
The dominant systematic errors come from the background model, lepton identification and
jet energy scale.

The complementary method~\cite{d0_lepton_jet_btag} 
uses a new Neural Network (NN) based $b$-jet tagging algorithm. The
chosen operating point has a $b$-jet tagging efficiency of $55\%$ and a fake tag rate of $1\%$. 
For the same fake tag rate this 
represents a $15\%$ increase in efficiency with respect to the previous $b$-jet tagging algorithm.
The cross-section result of 
$\sigma_{t\bar{t}} = 8.3^{+0.6}_{-0.5}{\rm(stat)}^{+0.9}_{-1.0}{\rm(syst)}\pm 0.5{\rm(lumi)~pb}$
is a combination from the NN-output in $8$ different channels,
considering $e$ and $\mu$ final states, $3$-jet events, or $4$ and more jet events,
$1$-tag or $2$-tags separately.
The jet-multiplicities for the combined $1$-tag and $2$-tags samples are shown in 
figure~\ref{fig:ljets}(middle and right respectively). 
Within the errors, the results are consistent between the kinematic and the b-tagging analyses.

CDF results in the $l$+jets channels have been presented at the previous Moriond QCD
conference~\cite{lisa} on an integrated luminosity of ${\cal L}=0.7~\rm{fb^{-1}}$. For
a method using only kinematic variables the result is
$\sigma_{t\bar{t}} = 6.0 \pm{0.6}{\rm(stat)}\pm{0.9}{\rm(syst)}\pm 0.3{\rm(lumi)~pb}$. The analysis using $b$-jet tagging yields to a cross-section of
$\sigma_{t\bar{t}} = 8.2 \pm{0.6}{\rm(stat)}\pm{0.9}{\rm(syst)}\pm 0.5{\rm(lumi)~pb}$.

\section{All jets channel}

Even though this channel has the highest branching ratio, it is largely 
dominated by multijet background.
The preselection of $ t\bar{t} \ra jets$ 
in the inclusive CDF multijet sample~\cite{cdf_all_jets}
requires  $6$ to $8$ jets with $p_T > 15~\rm{GeV}$ separated 
by $\Delta R > 0.5$. This preselection has a S/B ratio of 1/1300. 
To reduce the background a NN is used with $11$ kinematic input variables. 
A further improvement in the S/B ratio
is obtained by requiring a secondary vertex tag. The cross-section
is then measured from the number of observed tags with
an expectation of $n_{tag}=0.95 \pm 0.07$ per top event determined from Monte Carlo.
In total   $387$ signal and $846$ background events have been found for 
${\cal L} = 1.02~\rm{fb^{-1}}$.
The cross-section obtained is 
$\sigma_{t\bar{t}} = 8.3\pm{1.0}{\rm(stat)}^{+2.0}_{-1.5}{\rm(syst)}\pm 0.5{\rm(lumi)~pb}$
with the error on the jet energy calibration being the largest systematic contribution.

\section{Summary}
\begin{figure}
\begin{center}
\epsfig{figure=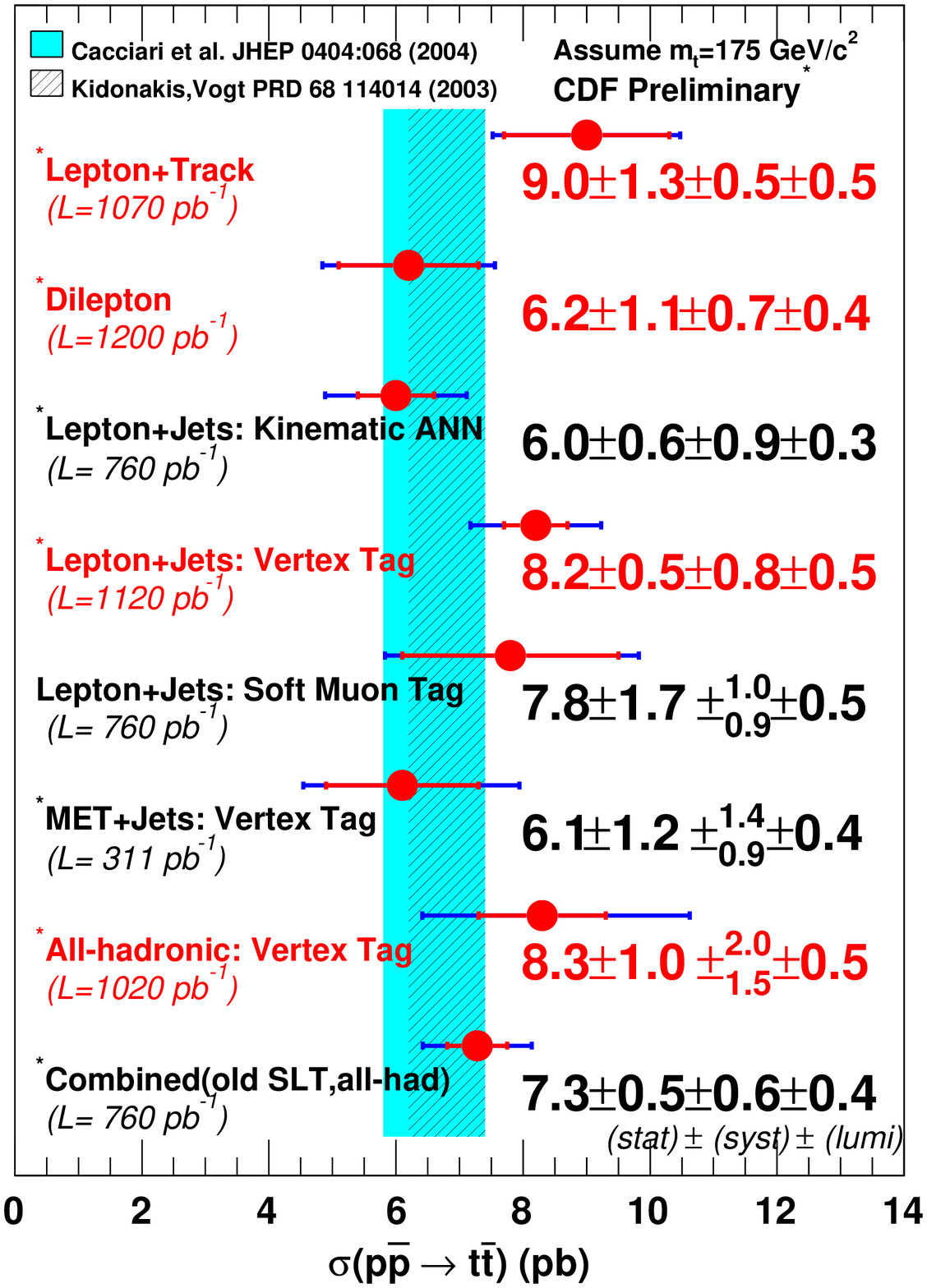,height=5cm}
\epsfig{figure=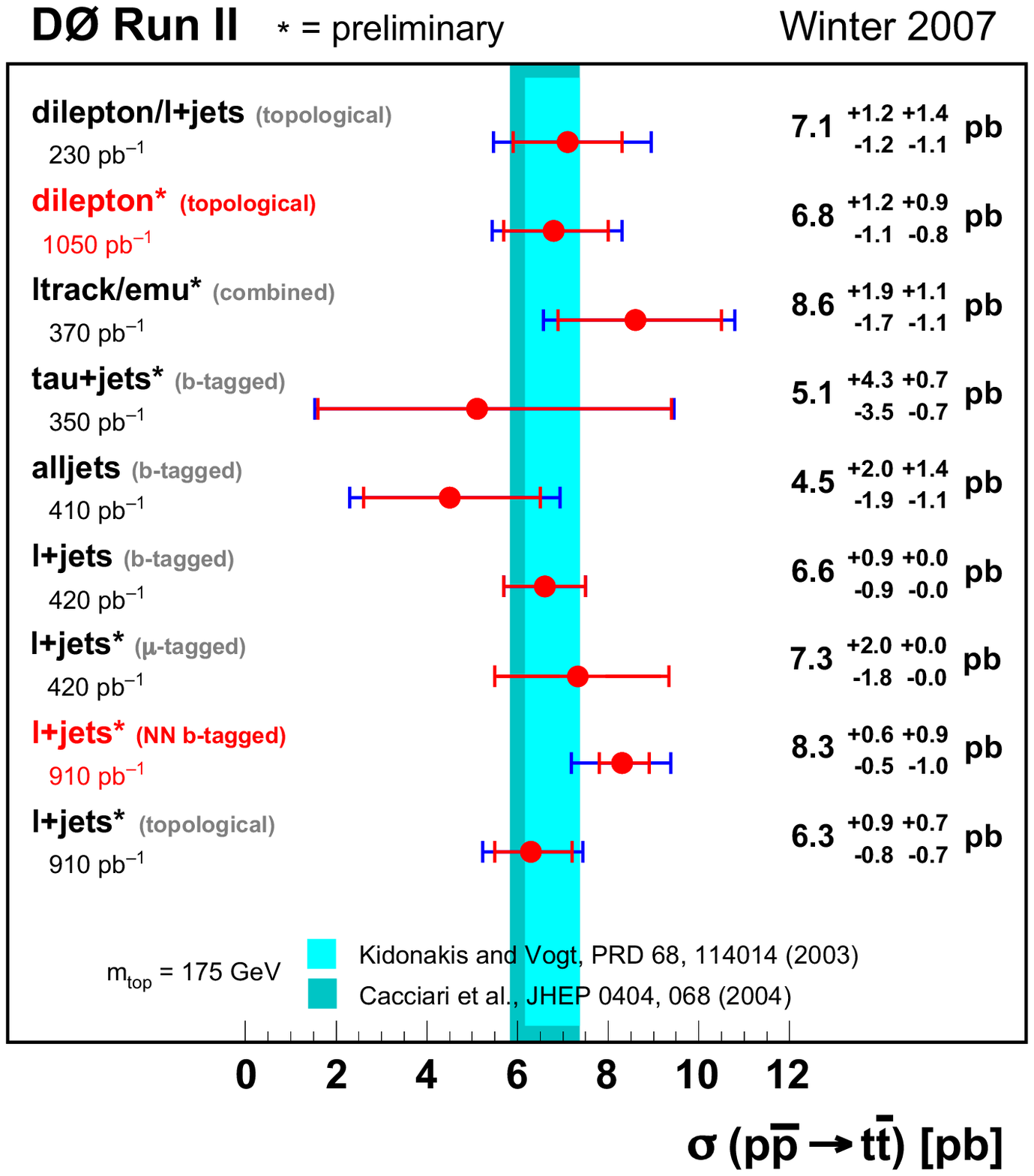,height=5cm}
\epsfig{figure=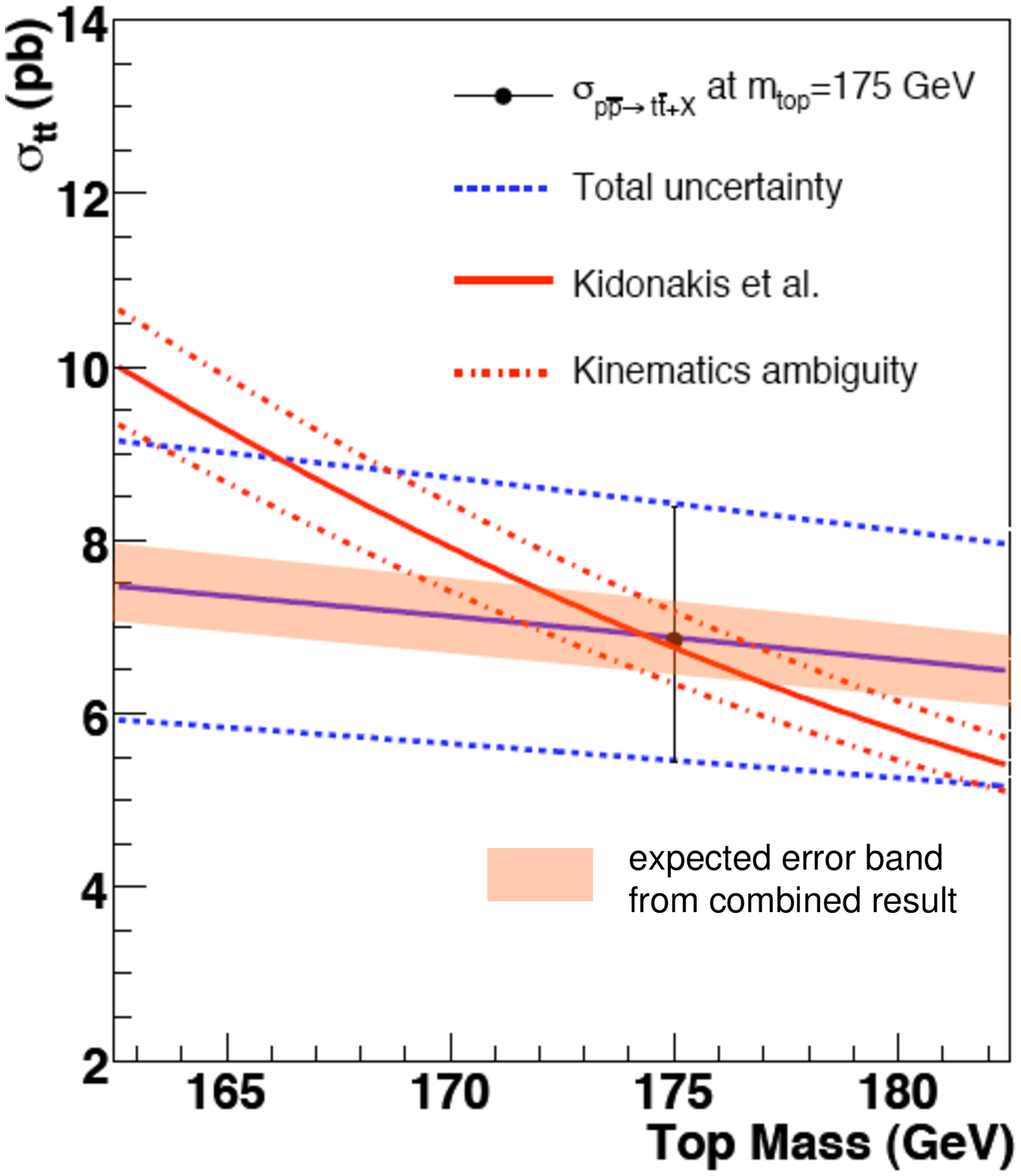,height=5cm, clip=}
\caption{Top quark pair production cross-section measurements: summary of the 
current CDF (left) results and D\O\ results (middle). The right figure shows the
$m_{\rm top}$ dependence of the D\O\ dilepton measurement.
\label{fig:summary}}
\end{center}
\end{figure}

Summaries of the CDF and D\O\ 
top quark pair production cross-section measurements 
are shown in figure~\ref{fig:summary}(left and middle respectively). 
All the measurements are in good agreement
with the NNLO-predictions.  
Results with an integrated luminosity of
about $1~\rm{fb^{-1}}$ are highlighted. 
With this luminosity the errors
have been sizably reduced and reach in some of the decay channels about $15\%$.
From a combination of all results an experimental error at the order of the 
theoretical error can be expected. 

During the conference the question was raised if $m_{\rm top}$ could be
determined from the cross-section measurements.
A first answer concerning the precision that could be achieved
can be interfered from figure~\ref{fig:summary}(right)
showing the dependence of the D\O\ dilepton cross-section
as a function of $m_{\rm top}$. The shaded band shows the assumption 
of the total error for a combined result of the currently measured current cross-sections 
to be of the size of the theoretical error and with the same $m_{\rm top}$ dependence than 
the di-lepton cross-section. 
A determination of $m_{\rm top}$ using the current
production cross-section would lead to an error on $m_{\rm top}$ of about $\pm 5~\rm{GeV}$.
With the full Run II statistics an experimental error half this size looks a reasonable guess.

\section*{Acknowledgments}
I would like to thank the CDF and D\O\ collaborations for presenting 
these results, in particular the top conveners
Kirsten Tollefson, Robin Erbacher, Elizabetha Shabalina, Ulrich Heintz, Michele Weber
for their help in preparing the talk, rehearsals and very useful comments, 
Kevin Lannon for a last minute plot and the organizers for another great Moriond!

\end{document}